\documentclass{aa}

\usepackage{CJKutf8}
\usepackage{graphicx}
\usepackage[breaklinks,colorlinks,citecolor=blue,linkcolor=blue]{hyperref}
\usepackage{txfonts}
\usepackage{lscape}
\usepackage{epstopdf}
\usepackage{CJK}
\usepackage{float}
\usepackage[absolute]{textpos}
\usepackage{amsmath}
\usepackage{mathrsfs}

\usepackage{arydshln}

\usepackage{cases}

\newcommand{\xunchuancnname}{\begin{CJK*}{UTF8}{gbsn}刘训川\end{CJK*}}

\begin{document} 

   \title{Turbulence in cascading: Origin of the variance and skewness of density function}

   \author{Xunchuan Liu (\xunchuancnname)
          }

   \institute{Leiden Observatory, Leiden University, P.O. Box 9513, 2300RA
Leiden, The Netherlands\\
              \email{liuxunchuan001@gmail.com}
             }


 
  \abstract{
Turbulent systems typically exhibit log-normal volume density probability density functions (PDFs) on the $s = \ln\rho$ scale, with their variance ($\sigma^2$) and skewness ($\tau$) empirically regulated by the Mach number ($\mathcal{M}$). In this work, we explain both the $\sigma^2$--$\mathcal{M}$ and $\tau$--$\mathcal{M}$ relationships from a thermodynamic and cascade perspective. Entropy conservation within the compressive modes yields the fiducial relation $\sigma^2 = \ln\left(1 + \mathcal{M}^2\right)$, while deviations from a monotonic energy cascade modify this law into a dilogarithm function. Although skewness exerts a negligible impact on the global $\sigma^2$--$\mathcal{M}$ scaling, the characteristic skewness parameter $\tau$ obeys a similar scaling law, $\tau \propto \ln\left(1 + \mathcal{M}^2\right)$. We demonstrate that the asymmetric wings of the global density PDF originate from underlying low-$s$ and high-$s$ skewed kernels. These kernels are governed by two distinct structural redistribution mechanisms—the mass-fraction and volume-fraction approaches—which exhibit a profound duality symmetry. Crucially, the high-$s$ skewed kernels are physically realizable and closely mirror the gravity-driven power-law  tails of column density PDFs observed in molecular clouds.
}


   \maketitle
%

\section{Introduction}
Turbulence is a ubiquitous phenomenon in fluid dynamics \citep{1941DoSSR..30..301K,BURGERS1948171} and is crucial for understanding a wide range of astrophysical and interstellar processes. Along with gravity and magnetic fields, turbulence is believed to govern the behavior of gas in molecular clouds \citep[e.g.,][]{2017A&A...599A..99O}, star-forming regions \citep[e.g.,][]{2025arXiv250117502L}, and various other astrophysical systems \citep[e.g.,][]{1999PPCF...41A.787V,2020GSL.....7...10R,2024A&A...684A.174P}. In these environments, it plays a pivotal role in regulating energy, momentum, and mass transport \citep[e.g.,][]{Larson81,2003ApJ...585..850M}. 
While the exact origin of interstellar turbulence remains fully unclarified, it is closely linked to the large-scale structure of the interstellar medium (ISM). This highly turbulent medium is driven by a combination of mechanisms, including shear forces \citep[e.g.,][]{2014PASJ...66...36M}, stellar feedback \citep[e.g.,][]{2009ApJ...695.1376C,2012PhDT.......284G}, cloud-cloud collisions \citep[e.g.,][]{2018PASJ...70S..57W}, and gravitational instabilities \citep[e.g.,][]{2015ApJ...814..131G}.

A key feature of turbulent flows in these systems is the probability density function (PDF) of the gas density ($\rho$), which has been found to follow a lognormal distribution \citep{1994ApJ...423..681V,1997MNRAS.288..145P,2010A&A...512A..81F}. Consequently, the logarithmic density, defined as $s=\ln(\rho)$, follows a normal distribution. The variance of this normal distribution (for $s$) is typically denoted as $\sigma^2$. In observations, directly measuring the three-dimensional volumetric density distribution is challenging because line-of-sight projection effects restrict measurements to the column density. Nevertheless, observations in regions not dominated by gravity reveal that the column density also follows a lognormal distribution \citep{2008MNRAS.390L..19B,2015A&A...575A..79S,2025arXiv250210897L}. To connect these quantities, numerical simulations have been heavily utilized; from these simulations, a fiducial empirical relationship has been established showing that $\sigma^2$ depends on the Mach number $\mathcal{M}$ via the form $\sigma^2 \sim \ln(1+b^2\mathcal{M}^2)$ \citep{1997MNRAS.288..145P,2010A&A...512A..81F,2012ApJ...761..149K,2013MNRAS.430.1880H}. Recent observations across various interstellar medium (ISM) environments, including diffuse atomic clouds \citep{2025arXiv250210897L} and dense molecular clouds \citep[e.g.,][]{2009ApJ...692...91G,2015ApJ...811L..28B}, also support this $\sigma^2$-$\mathcal{M}$ relation after accounting for line-of-sight integration effects \citep{2013ApJ...765...49G}. This strongly implies that the $\sigma^2$-$\mathcal{M}$ relation is a fundamental property governing compressible, isothermal turbulent gases.

However, despite its empirical success in simulations and broad application in observations, a complete physical and theoretical explanation for the $\sigma^2-\mathcal{M}$ relationship is still lacking. 
For instance, while the traditional $k$-$\epsilon$ turbulence model \citep{launder1979lectures, 2009MNRAS.394.1351P} successfully predicts empirical outcomes in engineered or terrestrial flows, it fails to provide a theoretical framework for the $\sigma^2-\mathcal{M}$ relation in isothermal, compressible systems typical of the ISM \citep{2001ApJ...557..736G,2003ApJ...586.1067H,2013MNRAS.430.1880H,2021RAA....21...24Y}. 
Alternative analytical approaches also face fundamental limitations. For highly supersonic regimes ($\mathcal{M} \gg 1$), modeling density jumps via one-dimensional shocks yields a scaling of $\sigma \sim \ln(\mathcal{M}^2)$ \citep{1998PhRvE..58.4501P}, which deviates from the standard empirical form. Conversely, treating the system as a cumulative effect of successive small density jumps naturally explains the lognormal distribution of $s$ via the central limit theorem, but the explicit dependence of $\sigma$ on $\mathcal{M}$ becomes obscured in the process \citep[e.g.,][]{1994ApJ...423..681V}.
Consequently, developing a self-consistent theory within a unified framework is highly desirable. Such a theory must simultaneously explain the $\sigma^2-\mathcal{M}$ relation across both subsonic and supersonic regimes, account for the distinct behaviors of different driving modes (solenoidal versus compressive) that only emerge in higher dimensions \citep{2010A&A...512A..81F,2012ApJ...761..149K}, and respect the cascading and random nature of the turbulent field. 
Furthermore, ideally, this framework should also address the observed deviations from the empirical $\sigma^2-\mathcal{M}$ relation at very high Mach numbers \citep{2008ApJ...688L..79F,2013MNRAS.430.1880H}.

Further more, numerical simulations demonstrate that for highly supersonic systems, particularly those driven by compressive forcing, the density PDFs remain lognormal at the high-density end but exhibit a distinct skewness toward the low-density (and low-$s$) regime \citep[e.g.,][]{2010A&A...512A..81F}. This low-$s$ skewness in the PDF has recently been observed in the \ion{H}{I} 21~cm line emission of a very high-velocity cloud using the Five-hundred-meter Aperture Spherical radio Telescope (FAST) \citep{2025NatAs...9.1366L}. 
However, the physical origin of this skewness (hereafter referred to as the skewness of turbulence) remains an open question. By interpreting turbulence as a cascading process that broadens the PDF across cascade levels \citep{1941DoSSR..30..301K,1996JPhy2...6..105C,PhysRevE.57.1737}, \citet{2013MNRAS.430.1880H} speculated that the low-$s$ skewness originates from the underlying convolution kernels associated with intermittency. Nevertheless, they did not explicitly demonstrate how intermittency generates these asymmetric tails from first physical principles. 
Thus, we also expect to 
provide a fundamental explanation for the skewness of density PDF, and 
explore possbile link between the skewness and the $\sigma^2-\mathcal{M}$ relation.

In this work, we provide a concise thermodynamic and cascade perspective on turbulence, establishing a unified theoretical framework to understand the $\sigma^2$--$\mathcal{M}$ relationship and PDF skewness across varying Mach numbers and driving modes. The paper is structured as follows: In Sect.~\ref{sec_backdef}, we introduce the essential background and definitions regarding density PDFs. In Sect.~\ref{sec_logpdf}, we provide a thermodynamic view of the $\sigma^2$--$\mathcal{M}$ relation. In Sect.~\ref{sec_frac}, we explore the origin of PDF skewness and its scaling relation to the Mach number. In Sect.~\ref{sec_discuss}, we discuss the physical scenarios of high-$s$ skewed PDFs along with several important considerations of this work. Finally, we summarize our main conclusions in Sect.~\ref{sec_summary}.

\section{Basic definitions}\label{sec_backdef}
Here, the term ``PDF'' denotes the probability density function of either the gas density $\rho$ or its logarithmic transformation $s = \ln(\rho)$. To ensure proper normalization, the density is rescaled such that the mean density $\langle \rho \rangle = 1$. A normalized PDF $f(s)$ must satisfy two fundamental normalization conditions:
\begin{align}
   & \int_{-\infty}^{\infty} f(s) \, ds = 1, \label{eq_normal_Psum} \\
   & \int_{-\infty}^{\infty} f(s) e^s \, ds = 1. \label{eq_normal_densitysum}
\end{align}
Let $g(s)$ and $h(s)$ be two normalized PDFs, and let $g * h$ denote their convolution. It is trivial that $g * h$ satisfies Eq.~\eqref{eq_normal_Psum}. It also follows that:
\begin{align}
     \int_{-\infty}^{\infty} (g * h)(s) e^s \, ds &= \int_{-\infty}^{\infty} \int_{-\infty}^{\infty} g(s - t) h(t) e^s \, dt \, ds \notag \\
     & = \int_{-\infty}^{\infty}  h(t)e^t \left(\int_{-\infty}^{\infty} g(s - t) e^{s-t} \, d(s-t)\right) dt \notag \\
     &= \int_{-\infty}^{\infty} h(t) e^t \, dt = 1.
\end{align}
Thus, the convolution $g * h$ is also normalized. Let $\sigma^2(g)$ denote the variance of $g(s)$, defined as:
\begin{equation}
    \sigma^2(g) = \int_{-\infty}^{\infty} g(s) (s-\langle s \rangle)^2 ds,
\end{equation}
where the mean logarithmic density is defined as:
\begin{equation}
    \langle s\rangle = \int_{-\infty}^{\infty} s\,g(s)\, ds. 
\end{equation}
The variance of a convolved distribution is strictly additive:
\begin{equation}
    \sigma^2(g * h) = \sigma^2(g) + \sigma^2(h).
\end{equation}

Next, consider the asymptotic behavior of PDFs with exponential tails. Let $g(x) \propto e^{ax}$ and $h(x) \propto e^{bx}$ for $x \to -\infty$, where the decay indices satisfy $a > b > 0$. In this regime, the convolution $(g * h)(x)$ evaluates to:
\begin{equation}
    (g * h)(x) \propto \int_{-\infty}^{x} e^{ay} e^{b(x - y)} \, dy \propto e^{bx} 
    \quad \text{as} \quad x \to -\infty. \label{eq_conv_tails}
\end{equation}
Therefore, when both kernels exhibit an exponential tail at the negative end, their convolution does not generate a new exponential scaling. Instead, the resulting distribution inherits the index of the component with the flattest, slowest-decaying tail.

Pioneering simulations in the 1990s \citep{1994ApJ...423..681V,1997MNRAS.288..145P} established that density PDFs in isothermal, compressible turbulent flows follow a lognormal distribution. Expressed in terms of the logarithmic density $s$, this distribution takes the standard form:
\begin{equation}
    f_{\rm normal}(s) = \frac{1}{\sqrt{2\pi}\sigma}\exp\left(-\frac{(s+\sigma^2/2)^2}{2\sigma^2}\right). \label{eq_lognormalfun}
\end{equation}
The mathematical coupling between the mean ($-\sigma^2/2$) and the variance ($\sigma^2$) arises directly from the mass conservation constraint given in Eq.~\eqref{eq_normal_densitysum}. 

\section{PDF variance from a thermodynamic perspective} 
\label{sec_logpdf}
First, we assume that the log-normal PDF form in Eq.~\eqref{eq_lognormalfun} is rigorous, independent of its underlying physical origin. Under this condition, we demonstrate from first principles that the empirical $\sigma^2$--$\mathcal{M}$ relation,
\begin{equation}
\sigma^2 \sim \ln\left(1 + b^2\mathcal{M}^2\right),
\label{sig_M_eq}
\end{equation}
can be derived, where the parameter $b$ depends on both the forcing mode and the Mach number regime. 

\subsection{Role of compressive mode}\label{pdf_basic}
A turbulent field can be decomposed into solenoidal and compressive components with a degrees-of-freedom ratio of 2:1. The Mach numbers of these two components are denoted as $\mathcal{M}_{\rm sol}$ and $\mathcal{M}_{\rm comp}$, respectively. 
In the subsonic regime ($\mathcal{M} \ll 1$), the compressive component degenerates into acoustic waves, scaling as $\sigma_{\rm comp} \sim \mathcal{M}_{\rm comp}$. Conversely, the solenoidal mode forms localized eddies where mass is compressed by centrifugal forces, yielding a scaling of $\sigma_{\rm sol} \sim \mathcal{M}_{\rm sol}^2$. If $\mathcal{M}_{\rm comp}$ and $\mathcal{M}_{\rm sol}$ scale proportionally to their degrees of freedom, $\sigma_{\rm comp}$ remains much larger than $\sigma_{\rm sol}$. However, because the compressive mode is highly prone to dissipation and radiative cooling, it becomes significantly weaker than the solenoidal mode while still dominating density fluctuations. Consequently, the Mach number in Eq.~\eqref{sig_M_eq} can be altered to yield:
\begin{equation}
\sigma^2 \sim \ln\left(1 + \mathcal{M}_{\rm comp}^2\right),
\label{sig_Mcomp_sub_eq}
\end{equation}
with $b = 1$ when evaluating the system purely in terms of $\mathcal{M}_{\rm comp}$.

In the supersonic regime ($\mathcal{M} \gg 1$), head-on collisions between opposing flows become the primary mechanism compressing the gas, allowing Eq.~\eqref{sig_Mcomp_sub_eq} to remain valid. When driven by compressive forcing, the turbulent system distributes energy evenly among the available degrees of freedom (at the same scale), which yields $b = \sqrt{1/3}$ in Eq.~\eqref{sig_M_eq} since $\mathcal{M}_{\rm comp}^2 \sim \frac{1}{3}\mathcal{M}^2$. Conversely, if the system is driven by solenoidal forcing, the compressive mode is not fully excited. We then consider a scenario where, at large scales, two-thirds of the total energy is stored in large-scale eddies, while the remaining one-third feeds into the energy cascade at inertial scales, behaving effectively like compressive driving. Under this scheme, we obtain $\mathcal{M}_{\rm comp}^2 \sim \frac{1}{9}\mathcal{M}^2$, which naturally yields $b = 1/3$, as verified by simulations \citep[e.g.,][]{2010A&A...512A..81F}.

Overall, these arguments indicate that $\mathcal{M}_{\rm comp}$ is the true governing parameter, meaning the traditional $\sigma^2$--$\mathcal{M}$ relation is fundamentally replaced by the $\sigma^2$--$\mathcal{M}_{\rm comp}$ formulation (Eq.~\ref{sig_Mcomp_sub_eq}) without requiring an auxiliary parameter $b$. Here, the fiducial relation is expressed as:
\begin{equation}
\sigma^2 \sim \ln\left(1 + \mathcal{M}_{\rm comp}^2\right).
\label{sig_Mcomp_sig_fiducal}
\end{equation}
In the following subsections, we simplify our notation by denoting $\mathcal{M}_{\rm comp}$ directly as $\mathcal{M}$ where no ambiguity arises, present a thermodynamic framework to derive Eq.~\ref{sig_Mcomp_sig_fiducal}, and explain the underlying mechanisms driving deviations from it.

\begin{figure}
    \centering
    \includegraphics[width=0.99\linewidth]{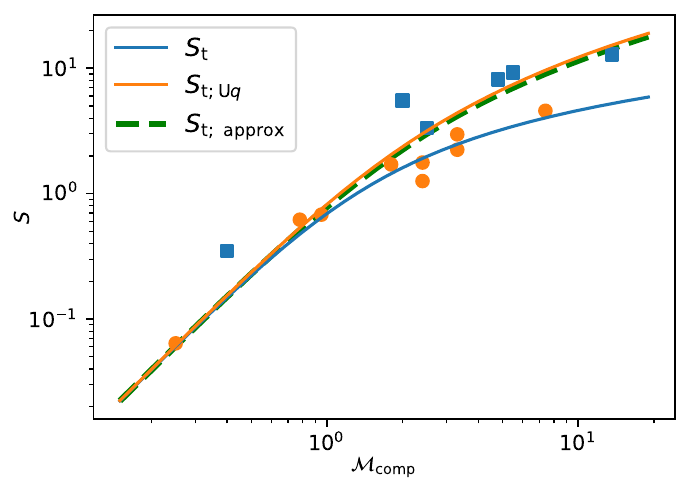}
    \caption{Comparison of the different functional forms of $S_{\rm t}$ (namely $S_{\rm t}$, $S_{{\rm t};\, {\rm U}q}$, and $S_{\rm t;\, approx}$), which serve as proxies for $\sigma^2$ (Sect.~\ref{sec_logpdf_basic}), from Eqs.~\ref{eq_st}, \ref{eq_st_uq}, and \ref{eq_st_approx} (Sect.~\ref{sec_logpdf}). Markers represent numerical simulation results tabulated by \citet{2013MNRAS.430.1880H} and references therein (excluding magnetically dominated cases). Yellow circles and blue squares denote simulations driven by solenoidal and compressive forcing modes, respectively. The horizontal axis indicates the Mach number of the compressive velocity component ($\mathcal{M}_{\rm comp}$, Sect.~\ref{pdf_basic}).
    }
    \label{fig:M-S_relation}
\end{figure}

\subsection{$\sigma^2-\mathcal{M}$ relation by entropy coupling}
\label{sec_logpdf_basic}
The development of a turbulent field represents an energy injection process spanning from the driving scale down to the dissipation scale. Treating the complete set of compressive modes within the system as a distinct sub-ensemble, the effective temperature of this sub-ensemble is defined as:
\begin{equation}
T_{\rm eff} = T_{\rm thermal} + T_{\rm turb},
\end{equation}
where the turbulent contribution is given by:
\begin{equation}
    T_{\rm turb} = \mathcal{M}^2 T_{\rm thermal}.
\end{equation}
During the initialization of the turbulent field, $T_{\rm eff}$ increases from $T_{\rm thermal}$ to $T_{\rm eff}$, with the turbulent kinetic energy cascading monotonically across scales down to the dissipation scale. The thermal entropy of the sub-ensemble increases by:
\begin{align}
    S_{\rm t} & = \int_{T_{\rm thermal}}^{T_{\rm eff}} \frac{1}{T}dT \label{eq_Sgeneral}\\
       &= \ln\left( \frac{T_{\rm eff}}{T_{\rm thermal}} \right)  \notag \\
       & = \ln\left(1+\mathcal{M}^2\right), \label{eq_st}
\end{align}
where the constant factor $\frac{1}{2}k_B$ is omitted for simplicity. Concurrently, the structural entropy of the established density field is defined via the PDF $f$ as:
\begin{align}
    S_{\rm s} &= -2\int \rho \ln(\rho) \, dV \notag\\
         &= -2\int f(\rho) \rho \ln(\rho) \, d\rho \notag\\
         &= -2\int f(s) s e^s \, ds. \label{eq_ss_s}
\end{align}
Here, the standard expression has also been rescaled by dividing by $\frac{1}{2}k_B$, ensuring consistency with Eq.~\eqref{eq_st}. For a purely lognormal density distribution given by Eq.~\eqref{eq_lognormalfun}, the structural entropy evaluates to:
\begin{align}
S_{\rm s} & = - 2\int_{-\infty}^{\infty} \frac{1}{\sqrt{2\pi}\sigma}\exp\left(-\frac{(s+\sigma^2/2)^2}{2\sigma^2}\right)  se^s ds\notag\\
           & = - 2\int_{-\infty}^{\infty} \frac{1}{\sqrt{2\pi}\sigma}\exp\left(-\frac{(s-\sigma^2/2)^2}{2\sigma^2}\right)  s ds\notag\\
           & = -2E[s'] = -\sigma^2, \label{eq_ss}
\end{align}
where $E[\cdot]$ denotes the expectation operator, and $s'$ follows a normal distribution with a mean of $\sigma^2/2$. 

Since the cascading process at scales larger than the dissipation scale is adiabatic and quasi-static---with the solenoidal mode acting as a lubrication mechanism that mitigates individual shock jumps (as considered by, e.g., \citealt{1998PhRvE..58.4501P})---the total entropy of the sub-ensemble is strictly conserved ($S = S_{\rm t} + S_{\rm s} = 0$). This equilibrium condition directly implies:
\begin{equation}
|S_{\rm t}| = |S_{\rm s}|, \label{eq_eqStSs}
\end{equation}
which immediately yields the fiducial relation in Eq.~\eqref{sig_Mcomp_sig_fiducal}.

\subsection{Deviation from the empirical $\sigma^2-\mathcal{M}$ relation}
\label{sec_delay_q}

For highly supersonic systems, the turbulent energy may not always cascade monotonically as assumed in Section~\ref{sec_logpdf_basic}. Instead, a fraction of the energy can bypass intermediate inertial scales and transfer directly below the dissipation scale via a discrete jump in wavenumber space. Here, we demonstrate how this jumping transfer influences the PDF variance.

\subsubsection{Variance as a dilogarithm function of $\mathcal{M}$}
To compensate for the jumping dissipation, which is prominently expected in turbulent systems driven by compressive forcing \citep{2010A&A...512A..81F,2011PhRvL.106q4502A}, we introduce a parameter-delay factor $q$ applied to $T_{\rm eff}$. This modifies Eq.~\eqref{eq_Sgeneral} to:
\begin{align}
    S_{{\rm t};\, q} &= \int_{0}^{T_{\rm turb}}\frac{1}{T_{\rm thermal}+qT} \, dT \\
    &= \frac{1}{q}\ln\left( 1+q\mathcal{M}^2 \right).
\end{align}
Reflecting the stochastic nature of fully developed turbulence, we treat $q$ as a random variable uniformly distributed between 0 and 1:
\begin{equation}
    q \sim U(0,1).
\end{equation}
Ensemble averaging over this distribution yields the net thermal entropy change:
\begin{align}
    S_{{\rm t};\, {\rm U}q} &= \int_0^1 S_{{\rm t};\, q} \, dq \notag\\
        &= \int_0^1 \frac{1}{q}\ln\left( 1+q\mathcal{M}^2 \right) \, dq \label{eq_st_uq}\\
        &= -{\rm Li}_2\left(-\mathcal{M}^2\right), \label{eq_st_uq_Li}
\end{align}
where ${\rm Li}_2$ denotes the dilogarithm function\footnote{${\rm Li}_2(x) = -\int_0^1 \frac{1}{q}\ln\left(1-qx\right) dq = -\int_{0}^{x} \frac{\ln(1-t)}{t} \, dt$.}. 

Equating $|S_{{\rm t};\, {\rm U}q}|$ and $|S_{\rm s}|$, as in Sect. \ref{sec_logpdf_basic}, yields:
\begin{equation}
    \sigma^2 = -{\rm Li}_2\left(-\mathcal{M}^2\right). \label{eq_sig_Li}
\end{equation}
Consequently, $\sigma^2$ is systematically larger than the value predicted by the fiducial relation (Eq.~\ref{sig_Mcomp_sig_fiducal}) in supersonic regimes.

\subsubsection{Comparison with simulations}
This deviation of $\sigma^2$ from the fiducial formula is well supported by numerical simulations of systems driven by compressive forcing \citep{2008ApJ...688L..79F,2009A&A...494..127S,2010A&A...512A..81F,2012ApJ...761..149K,2012MNRAS.423.2680M}. Notably, Eq.~\eqref{eq_sig_Li} matches the empirical functional form of $\sigma^2$ proposed by \citet{2013MNRAS.430.1880H} via data fitting (see Fig.~\ref{fig:M-S_relation}).
Because the dilogarithm function ${\rm Li}_2$ cannot be expressed in terms of elementary functions, we propose a straightforward analytical approximation. Its asymptotic behaviors scale as:
\begin{equation}
   S_{{\rm t};\, {\rm U}q} \sim \left\{
   \begin{array}{ll}
   \mathcal{M}^2 \sim \left[\ln(1+\mathcal{M})\right]^2 & \text{for } \mathcal{M} \ll 1, \\
   2\left[\ln(1+\mathcal{M})\right]^2 & \text{for } \mathcal{M} \gg 1.
   \end{array}
   \right.
\end{equation}
Thus, we propose the following analytical formula to approximate $S_{{\rm t};\, {\rm U}q}$, or equivalently, $-{\rm Li}_2\left(-\mathcal{M}^2\right)$:
\begin{equation}
    S_{{\rm t};\, {\rm U}q} \sim S_{\rm t;\, approx} = \left(2-e^{-\mathcal{M}}\right) \left[\ln(1+\mathcal{M})\right]^2.
    \label{eq_st_approx}
\end{equation}
This approximation $S_{\rm t;\, approx}$ deviates from the exact solution $S_{{\rm t};\, {\rm U}q}$ by less than six percent. A visual comparison between $S_{\rm t}$, $S_{{\rm t};\, {\rm U}q}$, and $S_{\rm t;\, approx}$ is provided in Fig.~\ref{fig:M-S_relation}.

Overall, we have demonstrated a physical derivation of the fiducial $\sigma^2-\mathcal{M}$ relation from a thermodynamic perspective. Furthermore, we show that stochastically driven deviations from a purely monotonic energy cascade modify this law into an dilogarithm function.

\section{PDF skewness and fraction strategies}\label{sec_frac}
Beyond the variance ($\sigma^2$) of the log-normal volume density PDF discussed in Sect.~\ref{sec_logpdf}, a critical structural feature is its skewness, or deviation from the standard log-normal functional form. Here, this skewness is characterized by an exponential tail in the PDF at the $s$ scale, which mathematically corresponds to a power-law tail at the $\rho$ scale. At the low-$s$ end, this skewness is not a universal feature, though it has been documented in certain numerical simulations and observational studies of purely turbulence-dominated systems \citep{2010A&A...512A..81F,2025NatAs...9.1366L}. Conversely, at the high-$s$ end, this skewness becomes particularly prominent in observations of molecular clouds where self-gravity takes effect \citep[e.g.,][]{2013ApJ...766L..17S,2014A&A...571A..95F,2015MNRAS.449.4465B,2017A&A...606L...2A}. Consequently, a comprehensive theoretical framework is required to explain precisely when and how such skewness emerges at the low-$s$ end from the perspective of turbulent cascading, while also establishing potential physical links between this turbulence-driven low-$s$ asymmetry and gravity-driven high-$s$ tails.

\subsection{PDF kernels in a truncated exponential shape}\label{sec_pdf_kernels}
By treating turbulence as a cascading process \citep{1941DoSSR..30..301K} accompanied by continuous density fragmentation during the energy cascade, we assume that this density redistribution exhibits self-similar behavior. Under this self-similarity assumption, a characteristic PDF kernel (in $s$), denoted as $f$, emerges at an individual elementary cascade step. The global density PDF can then be expressed as the successive convolution of a series of these structurally identical PDF kernels \citep{2006EPJB...51..229C,2013MNRAS.430.1880H}. Because exponential tails are preserved under the convolution operation (Eq.~\ref{eq_conv_tails}), the exponential tail observed in the global density PDF is highly likely an intrinsic characteristic of the underlying kernel. Through the successive convolution of $f$ over $n$ cascade steps, the resulting PDF, denoted as $f^{*n}$, approaches a normal distribution within the intermediate $s$ regime, whereas the exponential tails remain preserved at large $|s|$. Concurrently, its variance grows linearly as $\sigma^2(f^{*n}) = n\sigma^2(f)$, depending directly on both the variance of the individual kernel and the total number of cascade steps.

The simplest functional form of a skewed kernel is a truncated exponential function, as proposed by \citet{2013MNRAS.430.1880H} within the framework of intermittent turbulent cascades. This kernel is expressed as:
\begin{equation}
    f_{\tau;\ 0}(s) = \frac{1}{\tau} \exp\left( -\frac{s_{\rm cut}-s}{\tau} \right) (s < s_{\rm cut}), \label{eq_ftau0}
\end{equation}
where $\tau > 0$ quantifies the degree of skewness, and the $(s < s_{\rm cut})$ term acts as a truncation window function.
To ensure $\langle e^s \rangle = 1$, the truncation point $s_{\rm cut}$ must be explicitly related to $\tau$ via:
\begin{equation}
    s_{\rm cut} = \ln\left(1+\tau\right).
\end{equation}
However, the abrupt truncation of $f_{\tau;\ 0}$ at $s_{\rm cut}$ creates a discontinuity that is physically incompatible with the assumption of infinitesimal, continuous cascade steps. Another limitation of this formulation is that $f_{\tau;\ 0}$ in Eq.~\ref{eq_ftau0} cannot be physically extended to describe a high-$s$ skewed kernel. Motivated by these limitations, the following subsections introduce a modified functional form of the kernel, accompanied by a physical rationale that is compatible with both low-$s$ and high-$s$ skewed regimes.

\begin{figure}
    \centering
    \includegraphics[width=0.7\linewidth]{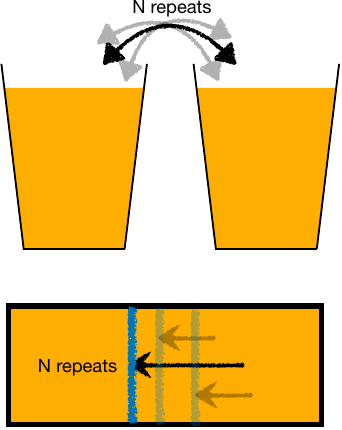}
    \caption{Schematic diagram of the different partition strategies: mass-fraction (upper) and volume-fraction (lower) strategies (Sect.~\ref{sec_frac}).}
    \label{fig:exchange}
\end{figure}

\begin{figure}[t]
    \centering
    \includegraphics[width=0.99\linewidth]{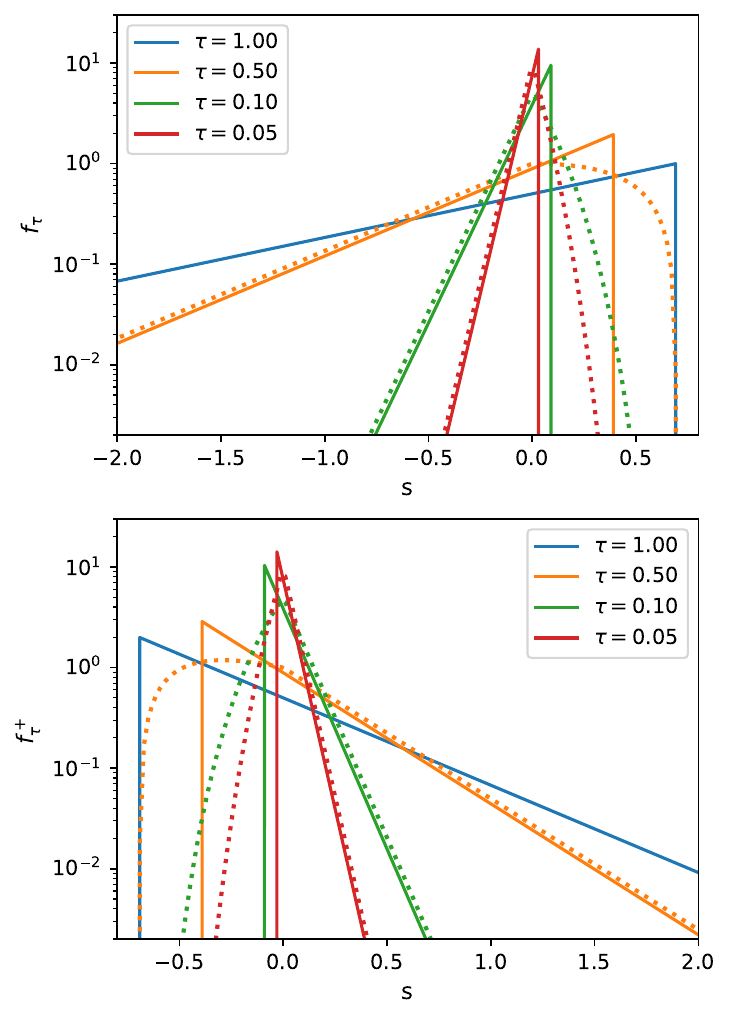}
    \caption{
   Comparison of the low-$s$ (upper panel) and high-$s$ (lower panel, marked by superscript $^+$) skewed PDF kernels (see Sect.~\ref{sec_frac}). 
   The solid and dotted lines represent the truncated exponential version ($f_{\tau;\ 0}$, Sects.~\ref{sec_pdf_kernels} and \ref{sec_fiduskk}) and the revised version ($f_\tau$, Sect.~\ref{sec_fraction_strategies}), respectively.
}
    \label{fig:ftau}
\end{figure}

\subsection{Modified PDF kernels via different partition strategies}\label{sec_fraction_strategies}
We adopt an energy (and consequently density) dichotomy strategy to describe the infinitesimal, continuous nature of an individual cascade step (Sect.~\ref{sec_frac}). Specifically, the mass from a parent cell at a given cascade level is dichotomously partitioned via a stochastic process into two subcells at the subsequent level. Under this framework, two distinct redistribution strategies naturally emerge: one where the subcells are constrained to maintain equal volumes (half the size of the parent cell), denoted as the mass-fraction strategy (Sect.~\ref{sect_massfrac}); and another where the volume of each subcell dynamically adjusts to drive density differentiation, denoted as the volume-fraction strategy (Sect.~\ref{sect_volume_frac}). As demonstrated below, these two strategies map directly onto the low-$s$ and high-$s$ skewed PDF kernels, respectively.

\subsubsection{Mass-fraction strategy}\label{sect_massfrac}
Here, we consider two subcells of equal unit volume, each initially containing a unit mass (upper panel of Fig.~\ref{fig:exchange}). During each trial mass-fraction process, the two subcells randomly exchange a mass increment \( \delta m \), where \( \delta m \) follows a uniform distribution between \(-1\) and \(1\). We assume that each individual cascade step consists of \(N\) independent virtual repetitions of this mass exchange process (see the discussion regarding $N$ in Sect.~\ref{sec_tau_M_empirical}). The specific mass exchange that minimizes the mass difference between the subcells (i.e., the realization yielding the smallest \(|\delta m|\)) is the one actually executed during that step\footnote{Mathematically, this strategy is analogous to an intermittent process that stochastically partitions the mass into $N+1$ cells, decoupling a single cell from the downstream cascade \citep[see, e.g.,][]{2013MNRAS.430.1880H}.}.

The dichotomous partition described above leads to an accumulated density distribution (CDF) for the subcell with the lower density ($\rho < 1$), given by:
\begin{equation}
    {\rm CDF}_{N,{\rm Mf}}(\rho) = \frac{1}{2} \rho^N. \label{eq_mf_masscdf}
\end{equation}
Consequently, the corresponding PDFs of \(\rho\) and \(s\) for this regime are expressed as:
\begin{align}
    f_{N,{\rm Mf}}(\rho) &= \frac{d \, {\rm CDF}_{N,{\rm Mf}}}{d\rho} = \frac{N}{2} \rho^{N-1},\\
    f_{N,{\rm Mf}}(s) &= f_{N,{\rm Mf}}(\rho) \frac{d\rho}{ds} = \frac{N}{2} \rho^N = \frac{N}{2} e^{Ns}. \label{eq_PDF_massstep}
\end{align}
A direct comparison between Eqs.~\ref{eq_PDF_massstep} and \ref{eq_ftau0} yields the parameter mapping \(\tau = 1/N\). For the counterpart subcell covering the density range \(1 < \rho < 2\), the local PDFs are governed by mass conservation constraints and are given by:
\begin{align}
    f_{N,{\rm Mf}}(\rho) &= f_{N,{\rm Mf}}(2-\rho) = \frac{N}{2} (2-\rho)^{N-1},\\
    f_{N,{\rm Mf}}(s) &= \frac{N}{2} (2 - e^s)^{N-1} e^s.
\end{align}
By transforming these piecewise solutions into the $\tau$ parameter space, we propose a modified kernel of the form:
\begin{equation}
    f_{\tau}(s) = \left\{
    \begin{aligned}
        &\frac{1}{2\tau} e^{s/\tau} && \text{for } s \leq 0, \\
        &\frac{1}{2\tau} \left( 2 - e^s \right)^{\frac{1-\tau}{\tau}} e^s
        && \text{for } 0 < s < \ln(2).
    \end{aligned}
    \right. \label{eq_ftau_modif}
\end{equation}
The modified kernel $f_{\tau}(s)$ preserves the low-$s$ exponential tail with a characteristic index of $1/\tau$, matching the behavior of the fiducial $f_{\tau;\ 0}$ (Eq.~\ref{eq_ftau0}), while introducing a smooth, continuously decreasing profile for $s > 0$ (see upper panel of Fig.~\ref{fig:ftau}). For small values of \(\tau\), the variance \( \sigma^2(f_{\tau}) \) deviates from the \(\tau^2\), the value for \( f_{\tau;\ 0} \). Instead, 
\begin{equation}
\sigma^2(f_{\tau}) \sim 2\tau^2 \quad \text{as } \tau \to 0.
\label{tau_modiff}
\end{equation}

\begin{figure}[!t]
    \centering
    \includegraphics[width=0.98\linewidth]{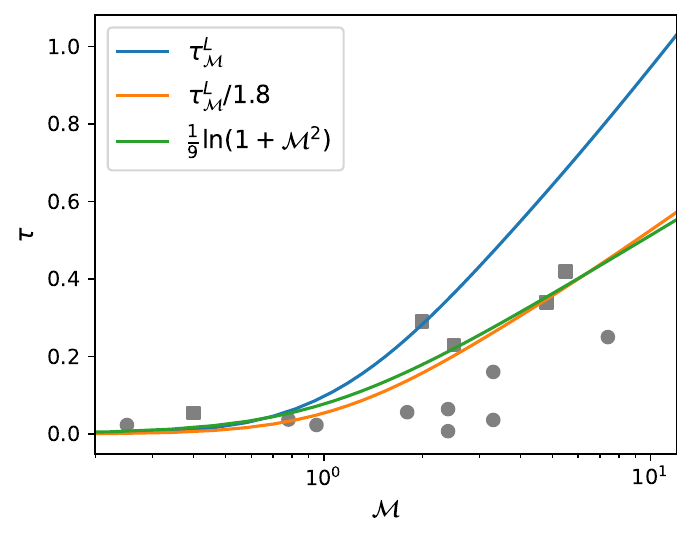}
\caption{Skewness parameters $\tau_\mathcal{M}^L$ and $\tau$ as functions of the Mach number $\mathcal{M}$ (see Sect.~\ref{sec_discuss_skewM} for details). Gray squares and circles denote the fitted $\tau$ values from numerical simulations driven by compressive and solenoidal forcing, respectively \citep{2013MNRAS.430.1880H}.}
\label{fig:tau_sigmadif}
\end{figure}

\subsubsection{Volume fraction and high-$s$ skewed PDF kernels}\label{sect_volume_frac}
In the volume-fraction strategy, the mass contained within each subcell is strictly conserved as unity, while the volume of each subcell is initially set to $\mathcal{V}=1$. Similarly to the mass-fraction framework, during each trial volume-fraction process, the two subcells randomly exchange a volume increment $\delta \mathcal{V}$, where $\delta \mathcal{V}$ is uniformly distributed. Again, we assume that each individual cascade step consists of \(N\) independent virtual repetitions of this volume exchange process. The specific volume exchange that minimizes the volume difference between the subcells is the one actually executed during that step. Under these conditions, the probability distribution of $\ln(\mathcal{V})$ obeys the functional form of $f_\tau$. Because the log-density scale maps as $s = \ln(1/\mathcal{V}) = -\ln(\mathcal{V})$, the temporary unweighted PDF can be written as:
\begin{equation}
    f_{\tau;\, {\rm tmp}}(s) = f_\tau(-s).
\end{equation}
Crucially, because the volume of the subcells is variable rather than invariant, we must apply a volume-weighting factor to the PDF kernel to account for the changing geometric sizes. This yields the final volume-weighted kernel:
\begin{equation}
    f_{\tau}^+(s) = e^{-s} f_{\tau;\, {\rm tmp}}(s) = e^{-s} f_\tau(-s).
\end{equation}
Note that $f_{\tau}^+={\rm Dual}[f]$ (Sect. \ref{sec_dualpdf}) is normalized.

\subsubsection{Symmetry between the low-$s$ and high-$s$ regimes}
\label{sec_symmetry} \label{sec_dualpdf}
Let \( f(s) \) be any normalized PDF satisfying Eqs.~\ref{eq_normal_Psum} and \ref{eq_normal_densitysum}. We define the dual transformation of \( f \) as:
\begin{equation}
f^{\rm dual}(s) \equiv \text{Dual}[f](s) = e^{-s} f(-s). \label{eq_dual_define}
\end{equation}
Consequently, $f^{\rm dual}$ satisfies the following integral identities:
\begin{align}
    &\int_{-\infty}^{\infty} f^{\rm dual}(s) \, ds = \int_{-\infty}^{\infty} e^{-s} f(-s) \, ds = \int_{-\infty}^{\infty} e^{s} f(s) \, ds = 1, \\
    &\int_{-\infty}^{\infty} e^s f^{\rm dual}(s) \, ds = \int_{-\infty}^{\infty} f(-s) \, ds = \int_{-\infty}^{\infty} f(s) \, ds = 1,
\end{align}
which confirms that ${\rm Dual}[f]$ remains a valid, normalized PDF. Furthermore, the dual operation commutes with the convolution operator, as demonstrated by the identity:
\begin{align}
    f^{\rm dual}(s) * g^{\rm dual}(s) &= \left[f(-s)e^{-s}\right] * \left[g(-s)e^{-s}\right] \notag \\
    &= \left[f(-s) * g(-s)\right] \cdot e^{-s} \notag \\
    &= (f * g)(-s) \cdot e^{-s} = (f * g)^{\rm dual}(s). \label{eq_fg_conv_dual}
\end{align}
Assuming that all low-$s$ and high-$s$ skewed PDFs are generated via the successive convolution of a series of kernels $f_\tau$ and $f_\tau^+$, respectively, these skewed distributions form a closed space under the dual operation. It implies a profound symmetry between the low-$s$ and high-$s$ regimes, strongly suggesting that their underlying physical mechanisms can be encapsulated within a unified framework.

\subsection{Fiducial skewed PDF kernels} \label{sec_fiduskk}
For mathematical tractability, the fiducial low-$s$ skewed kernel is chosen as the truncated exponential kernel $f_{\tau;\ 0}$ (see Eq.~\eqref{eq_ftau0} in Sect.~\ref{sec_pdf_kernels}). Inspired by the fractional frameworks outlined in Sect.~\ref{sec_fraction_strategies}, the fiducial high-$s$ skewed kernel, $f_{\tau;\ 0}^+(s)$, is formulated as the dual transformation (Sect. \ref{sec_dualpdf}) of $f_{\tau;\ 0}$:
\begin{equation}
    f_{\tau;\ 0}^+(s) = \mathrm{Dual}[f_{\tau;\ 0}](s) = \frac{1}{\tau} \exp\left( -\frac{s_{\rm cut}+s}{\tau} \right) e^{-s} (s > -s_{\rm cut}).
\end{equation}
The, all low-$s$ and high-$s$ skewed PDFs are generated via the successive convolution of a series of kernels $f_{\tau;\ 0}$ and $f_{\tau;\ 0}^+$.

\subsection{$\tau$-induced variance broadening}\label{sec_discuss_skewM}
The $\sigma^2-\mathcal{M}$ relation relies directly on the mathematical connection between the variance and the entropy-like quantity $S_s$ of a given PDF. For a fiducial low-$s$ skewed kernel, $f_{\tau;\ 0}$, the variance scales simply as:
\begin{equation}
    \sigma^2(f_{\tau;\ 0}) = \tau^2.
\end{equation}
Concurrently, the corresponding structural integral $S_s(f_{\tau;\ 0})$ can be derived compactly by exploiting the duality symmetry as follows:
\begin{align}
     S_s(f_{\tau;\ 0}) &= -2\int_{-\infty}^{\infty} f_{\tau;\ 0}(s) s e^s \, ds = 2\int_{-\infty}^{\infty} f_{\tau;\ 0}(-s) s e^{-s} \, ds \notag \\
     &= 2\int_{-\infty}^{\infty} f_{\tau;\ 0}^+(s) s \, ds = \frac{2}{1+1/\tau} - 2\ln(1+\tau) \notag \\
     &= -\sigma^2(f_{\tau;\ 0}) + \mathcal{O}(\tau^3).
\end{align}
That is for small $\tau$, 
\begin{align}
    |S_{\rm s}(f_\tau)| \sim \sigma^2(f_{\tau;\ 0}) 
                  \sim \tau^2 \quad {\rm as}\ \tau \to 0.
\end{align}
The ratio 
\begin{equation}
 X(f_{\tau;\ 0}) \equiv \sigma^2 / |S_{\rm s}|   
\end{equation}
grows monotonically as $\tau$ increases. 
Note that by $n$-repeat subsessive convolution, $f_{\tau;\ 0}^{*n}(s)$ approaches
normal distribution, and  $X(f_{\tau;\ 0}^{*n})\to 1$ for large $n$.

For a given PDF $f$, Eq.~\ref{eq_eqStSs} yields:
\begin{equation}
    \sigma^2(f) = X(f) \ln\left(1+\mathcal{M}^2\right). \label{eq_XM}
\end{equation}
Since $X$ is greater than unity for a skewed PDF, Eq.~\ref{eq_XM} may provide an alternative physical rationale for the broadening of the variance ($S_{\rm t;\, approx}/S_{\rm t}$) beyond the standard fiducial relation (Eq.~\ref{sig_Mcomp_sig_fiducal}). By equating 
\begin{equation}
    X(f_{\tau;\ 0}) = \frac{S_{\rm t;\, approx}}{S_{\rm t}},
\end{equation}
a characteristic skewness parameter $\tau$ can be solved (denoted as $\tau_\mathcal{M}^L$). This solved value can be approximated (see Fig.~\ref{fig:tau_sigmadif}) as:
\begin{equation}
  \tau_\mathcal{M}^L \sim \frac{1}{5} \ln\left(1+\mathcal{M}^2\right),
\end{equation}
represents the lower limit of $\tau$ required to account for the observed variance broadening across all possible low-$s$ skewed PDFs.
As shown in Figure \ref{fig:tau_sigmadif}, $\tau_\mathcal{M}^L$ is always larger than the simulated values for turbulent systems driven by compressive forcing \citep{2013MNRAS.430.1880H}. 
This discrepancy implies that PDF skewness alone cannot fully account for the observed variance broadening, confirming that an alternative physical mechanism (such as the parameter-delay framework introduced in Sect.~\ref{sec_delay_q}) is indeed required.

\subsection{$\tau-\mathcal{M}$ relation}\label{sec_tau_M_empirical}
Further comparison between $\tau_\mathcal{M}^L$ and simulated values (for systems driven by compressive forcing) reveals that:
\begin{equation}
\tau_\mathcal{M} \sim \frac{1}{1.8}\tau_{\mathcal{M}}^L \sim \frac{1}{9}\ln\left(1+\mathcal{M}^2\right). \label{eq_tMeq}
\end{equation}
This relation implies that, although skewness does not drive a significant broadening of the PDF variance, we can conversely estimate the skewness parameter $\tau$ directly from the PDF variance. Consequently, the number of independent virtual trials required for each cascade step (Sect.~\ref{sec_fraction_strategies}) scales as:
\begin{equation}
    N \sim \frac{1}{\tau_\mathcal{M}} \sim \frac{9}{\ln\left(1+\mathcal{M}^2\right)}.
    \label{NMrela}
\end{equation}
The coefficient $9$ appearing in the denominator of Eq.~\ref{eq_tMeq} (and correspondingly in the numerator of Eq.~\ref{NMrela}) may be interpreted as $d^2$, where $d=3$ is the spatial dimension of the system, though a physical justification remains outside the scope of this work. This coefficient is expected to be replaced by a significantly larger value for systems driven by solenoidal forcing. This scaling modification shifts the distribution toward higher values of $N$, which naturally suppresses the asymmetry and yields nearly symmetric or only slightly skewed PDFs.

In the subsonic regime where $\mathcal{M} \ll 1$, a characteristic turbulent eddy exhibits a density fluctuation amplitude of $\delta \rho \propto \mathcal{M}^2$, which is systematically decohered by convective nonlinear terms over a dynamic timescale of $\mathcal{O}\left(\mathcal{M}^{-2}\right)$ eddy turnover times
\citep[e.g.,][]{1992PhFlA...4..148G}. Statistically, this implies that an eddy undergoes a complete binary fragmentation event only after surviving for $\propto \mathcal{M}^{-2}$ dynamic periods. This scaling matches the behavior of Eq.~\ref{NMrela}, given that $\ln\left(1+\mathcal{M}^2\right) = \mathcal{M}^{-2}$ as $\mathcal{M} \to 0$. This alignment strongly corroborates the physical validity of the analytical framework developed in this work, which unifies the $\sigma^2-\mathcal{M}$ relation and the $\tau-\mathcal{M}$ relation across a broad range of Mach numbers.




\section{Discussion}\label{sec_discuss}


\subsection{High-$s$ skewed systems}\label{sec_starformation}
The low-$s$ and high-$s$ skewed PDF kernels from two distinct density partition strategies are dual to each other, exhibiting a highly symmetric mathematical architecture (Sect.~\ref{sec_fraction_strategies}). Thus, the high-$s$ skewed kernel is likely physically realizable in isothermal turbulent systems, even without an explicit gravitational term.

The momentum equation for an isothermal fluid is:
\begin{equation}
\frac{\partial \mathbf{v}}{\partial t} + (\mathbf{v} \cdot \nabla) \mathbf{v} = -\frac{1}{\rho} \nabla P - \nabla \phi,
\end{equation}
where the standard equation of state satisfies:
\begin{equation}
    P = \rho c_s^2 = \rho T_{\rm thermal}. \label{eq_normalpressure}
\end{equation}
Here, $c_s$ is the isothermal sound speed, $c_s^2 = T_{\rm thermal}$ omitting constants, and $\phi$ is the gravitational potential. 

In a coupled turbulence-gravity system, a gravity-bound region compressed by external turbulence initially provides positive thermal pressure due to its dense boundaries. However, as the internal structure adjusts under self-gravity, this region exerts a weakened or negligible thermal pressure to counteract external compression. To capture this behavior analogously, we construct a toy system excluding the gravitational force term $\nabla \phi$, introducing instead a modified, non-local pressure:
\begin{equation}
    P = \rho T_{\rm thermal} \cdot \mathcal{L}\left( \mathcal{D}, \det(\mathcal{H}) \right).
\end{equation}
Here, $\mathcal{L}$ is a modulation function dependent on the mass flux divergence $\mathcal{D} = \nabla \cdot (\rho \mathbf{v})$ and the determinant of the spatial density Hessian matrix, $\det(\mathcal{H})$. The simplest piecewise form of $\mathcal{L}(x,y)$ is:
\begin{equation}
    \mathcal{L}(x,y) = \left\{ 
    \begin{aligned}
        &1 - \Omega \quad && \text{for } x > 0 \text{ and } y < 0, \\
        &1 \quad && \text{otherwise,} 
    \end{aligned}
    \right. \label{eq_L_modi}
\end{equation}
where $x = \mathcal{D}$ and $y = \det(\mathcal{H})$. Under this formulation, when gas undergoes local compression ($x \leq 0$), the pressure behaves normally (Eq.~\ref{eq_normalpressure}). Conversely, when a density peak ($y < 0$) begins to expand or rebound ($x > 0$), the thermal pressure is reduced by a factor of $1-\Omega$. 
If $\Omega = 0$, the system reverts to its standard isothermal state. If $\Omega = 1$, a convergent region normally compresses, but the resulting density enhancement can never spontaneously relax or expand back. This asymmetric, irreversible trapping encapsulates the volume-fraction strategy. 


We hypothesize that within this framework, $\Omega$ quantifies the efficiency of the volume-fraction cascade mechanism. Under true gravity, molecular clouds above a specific density threshold display analogous behavior. This yields characteristic high-$s$ skewed PDFs before global gravitational collapse. Investigating the numerical behavior of this modified closure (Eq.~\ref{eq_L_modi}) and its mapping onto empirical cloud PDFs offers a promising diagnostic tool for future hydrodynamical modeling.

\subsection{Considerations and limitations}\label{sect_short}
Several key limitations and physical considerations apply to the current framework. First, the modified pressure closure introduced in Sect.~\ref{sec_starformation} differs fundamentally from the classic pressureless Burgers fluid \citep{BURGERS1948171}. Even in the extreme limit where $\Omega = 1$, our formulation prevents unphysical infinite density singularities because standard thermal pressure remains fully active during active compression phases. At present, this modified fluid framework remains a conceptual toy model; systematic numerical simulations are required to evaluate the explicit morphology and statistical stationarity of its resulting density structures.

Second, this framework neglects the effects of magnetic fields, which crucially regulate interstellar medium (ISM) turbulence. Extending our thermodynamic cascade perspective to magnetohydrodynamics (MHD) introduces severe structural complications. In the strong-field regime, the plasma dynamics effectively constrain the flow to a two-dimensional (2D) anisotropy. Conversely, in the presence of mild or weak magnetic fields, the turbulence decomposes into a complex superposition of Alfvén, fast, and slow MHD waves operating at distinct characteristic velocities \citep[e.g.,][]{BeresnyakLazarian+2019}. While fast magnetoacoustic waves propagate analogously to sound waves, slow waves introduce a highly intricate, non-linear coupling between density fluctuations and the local magnetic topology. For instance, magneto-hydrostatic structures where overdense filaments are confined by magnetic pressure gradients represent a specialized degenerate limit of slow-mode configurations. Consequently, it remains an open question whether the net entropy changes driven by turbulent cascade dissipation and structural fragmentation remain coupled via the simple, direct relations derived here for pure hydrodynamic systems.


\section{Summary}\label{sec_summary}
In this work, we present a thermodynamic and cascading theory to explain the origins of the variance and skewness of density PDFs (on the $s=\ln\rho$ scale) for isothermal compressible gas under turbulence. The main results include:

\begin{itemize}
    \item[1.]We obtain the empirical relation \(\sigma^2 = \ln\left(1 + \mathcal{M}^2\right)\) from entropy conservation (thermal entropy plus structural entropy) within the compressive modes. Further, we introduce a delay parameter to compensate for the deviation from monotonic cascading, leading $\sigma^2$ to be a dilogarithm function of $\mathcal{M}$, as confirmed by literature simulations.
    \item[2.]The asymmetric wings of the density PDF originate from underlying low-$s$ and high-$s$ skewed PDF kernels. These kernels are governed by two distinct structural redistribution mechanisms—the mass-fraction and volume-fraction approaches, respectively—during energy cascading.
    The skewed PDF kernels exhibit highly symmetric structures. 
    \item[3.] Although skewness exerts a negligible impact on the global $\sigma^2$--$\mathcal{M}$ scaling, the characteristic skewness parameter $\tau$ obeys a similar scaling law proportional to $\ln\left(1 + \mathcal{M}^2\right)$. This indicates that PDF skewness and variance are two relevant, but not fully dependent, statistical parameters of turbulent systems.

    \item[5.] We speculate that the high-density power-law tails of the column density PDFs of molecular clouds may be related to the high-$s$-skewed PDF kernels. Inspired by this, we briefly describe a form of turbulent system that may prefer the volume-fraction strategy.
\end{itemize}

Overall, this framework provides a physical baseline for interpreting the statistical properties of density PDFs in compressible turbulent flows. While these analytical results offer valuable insights, systematic numerical simulations remain essential to fully evaluate the structural complexities of turbulence, particularly under the anisotropic influence of interstellar magnetic fields, which remain unexplored here.

\begin{acknowledgement}
X. Liu acknowledges the support of the Strategic Priority Research Program of the Chinese Academy of Sciences  under Grant No. XDB0800303, and the National Key R\&D Program of China under Grant No. 2022YFA1603100. 
\end{acknowledgement}










\bibliography{nPDF_paper_merge}
\bibliographystyle{aa}

\end{document}